\documentclass[preprint]{aastex}

\slugcomment{Accepted for publication in ApJ}

\shorttitle{Dynamics of the outer parts of $\omega$ Centauri}

\shortauthors{Da Costa}

\begin{document}

\title{The dynamics of the outer parts of $\omega$ Centauri}

\author{G. S. Da Costa}

\affil{Research School of Astronomy \& Astrophysics, Australian 
National 
University, Mt~Stromlo Observatory, via Cotter Rd, Weston, ACT 2611,
Australia}

\begin{abstract}
The multi-object fibre-fed spectrograph AAOmega at the Anglo-Australian Telescope has been
used to establish and measure accurate ($\leq$1 km s$^{-1}$) radial velocities for a new sample of 
members in the outer parts of the stellar system $\omega$ Centauri.  The new sample more
than doubles the number of known members with precise velocities that lie between 25$\arcmin$
and 45$\arcmin$ from the cluster center.  Combining this sample with earlier work confirms that 
the line-of-sight
velocity dispersion of $\omega$ Cen remains approximately constant at $\sim$6.5 km s$^{-1}$ in the
outer parts of the cluster, which contain only a small fraction of the total cluster stellar mass.   It is
argued that the approximately constant velocity dispersion in the outer regions is most likely a 
consequence of external influences, 
such as the tidal shock heating that occurs each time $\omega$ Cen crosses the Galactic plane.
There is therefore no {\it requirement} to invoke dark matter or non-standard gravitational theories.
\end{abstract}

\keywords{globular clusters: general; globular clusters: individual ($\omega$ Centauri,
NGC~5139); stars: kinematics and dynamics}

\section{Introduction}

The stellar system $\omega$ Centauri has been known to be unusual, at least as regards its
stellar population, for almost four decades.  There is now an extensive body of work which shows
that, unlike the situation for most globular clusters, the member stars of $\omega$ Cen possess 
a large range in heavy element abundance together with distinctive element-to-iron abundance
ratios \citep[e.g.,][and the references therein]{JP10}.  A substantial spread in Helium abundance
has also been inferred from the observed abundances and structure of the lower main sequence 
in the cluster
color-magnitude diagram \citep[e.g.,][]{JEN04, GP05}.  Investigations of the metallicities of 
$\omega$ Cen stars in the vicinity of the main sequence turnoff also suggest that the cluster has an
age spread of perhaps 2 Gyr \citep[e.g.,][and the references therein]{EP11}.  
Together these characteristics have led to
the suggestion that $\omega$ Cen has not evolved in isolation but is instead the nuclear remnant of
a now disrupted nucleated dwarf galaxy that was accreted by the Milky Way \citep[e.g.,][]{KCF93}.
\citet{BF03} have shown that despite the tightly bound and retrograde current orbit of $\omega$~Cen,
such a disruption and accretion process is dynamically plausible.  Nevertheless, the spectroscopic
survey of \citet{DC08} showed that there is little evidence for any significant extra-tidal population
surrounding $\omega$ Cen at the present day, consistent with the photometric study of
\citet{DL03}.  \citet[][hereafter DC08]{DC08} give an upper 
limit of 0.7\% for the fraction
of the cluster mass contained between 1 and 2 cluster tidal radii.  This result requires the tidal stripping
and disruption process of the postulated progenitor system to be largely complete at early epochs with
the stars from the disrupted dwarf galaxy now widely distributed around the Galaxy \citep[e.g.,][]{WDB10,
SJM12}.

While the nucleosynthetic history of $\omega$ Cen is complicated and not fully understood, the
dynamics of the present-day stellar system, at least for the part of the cluster containing most of the
stellar mass, are relatively well established.  There have been a number of models of the system
including those of \citet{GM87}, \citet{MM95}, \citet{DM97}, \citet{GH03} and \citet{vMA10}, 
all of which, within their adopted assumptions, reproduce well the available observational data.
The most detailed model is that of \citet{vdV06}.  This axisymmetric dynamical model, which includes 
rotation and radially varying anisotropy, suggests that the mass-to-light ratio of $\omega$ Cen does not
change with radius -- the variation in the model $M/L_{V}$ value  does not deviate
significantly from the best-fit constant value of 2.5 (solar units) out to the limits of the modelled data
at $r$ $\sim$ 20$\arcmin$ \citep{vdV06}.

However, it is necessary to keep in mind that models such as that of \citet{vdV06} are constrained by
the extent of available observational data.  In the case of the velocity dispersion profile for 
$\omega$~Cen, the
data have been limited, until relatively recently, to a radius of approximately 20$\arcmin$
from the cluster center.  While this radius ($\sim$4 half-light radii) contains most of the cluster stellar
mass, it nevertheless is less than half the nominal  ``tidal radius'' of $\omega$~Cen 
(57$\arcmin$; see the discussion in DC08).  The lack of information on the velocity dispersion 
profile at large radii may mean we are currently missing some interesting astrophysics.  For example, 
if $\omega$~Cen is the nuclear remnant of a disrupted dwarf galaxy then it is possible it has retained
some of the dark matter content of the original system.  One of the best places to constrain the dark matter
content is in the outer parts of the cluster where the stellar densities are low \citep[e.g.,][]{CL00,MS05}.

\citet{SM03} presented the first data for the line-of-sight velocity dispersion of $\omega$~Cen beyond
$\sim$20$\arcmin$ from the cluster center.  They used accurate radial velocities for 75 members 
with  $\sim$20 $\leq$ $r\arcmin$ $\leq$ 30 to show that the cluster velocity dispersion profile 
may be relatively flat
beyond 20$\arcmin$.  This is in contrast to the monotonically declining dispersion profile
expected for a system in dynamical equilibrium in
which mass follows light.  \citet{SM03} chose to interpret their results as indicating the breakdown of
Newtonian dynamics in a weak acceleration regime.  However, this interpretation has been 
questioned by, for example, \citet{BGK05} who argue that the external influence of the Milky Way
on clusters such as $\omega$~Cen, that lie relatively close the Galactic Center, is sufficiently large
that the effective acceleration is larger than the critical MOND constant $a_{0}$; thus Newtonian dynamics
should still apply. 

Two additional studies of the velocity dispersion in the outskirts of $\omega$~Cen have recently 
appeared.  In the first, \citet{So09} conducted a survey for new $\omega$~Cen members in the outer
regions of the cluster and
combined their radial velocity results with those from the earlier study of \citet{EP07} to generate
a velocity dispersion profile for the cluster that reached a radial distance of $\sim$32$\arcmin$.  The
typical uncertainty in the velocity dispersion measures was $\leq$1 km s$^{-1}$.  \citet{So09} claim
that the velocity profile decreases monotonically from the center outwards though their outermost
data point lies above the previous point by more than the combined (1$\sigma$) errors.   \citet{So09}
note that this occurence might be compatible with the onset of tidal heating in the outskirts of the cluster.
Nevertheless, the \citet{So09} data are not inconsistent with a constant velocity dispersion beyond $r$
$\approx$ 20$\arcmin$.   The \citet{So09} sample contains 98 $\omega$~Cen members beyond
$r$ $\approx$ 20$\arcmin$ but of these stars only 13 lie beyond 30$\arcmin$.

The second recent paper is that of \citet{SF10} in which the data sets of \citet{So09} and \citet{SM03}
have been combined to provide a further estimate of the velocity dispersion profile. 
The addition of the \citet{SM03} velocities increases the number of 
stars with accurate velocities for radial distances between 20$\arcmin$ and 30$\arcmin$ but does
not contribute any new members beyond $r$ $\approx$ 30$\arcmin$.  The combined data are
consistent with a flattening of the velocity dispersion beyond $r$ $\approx$ 20$\arcmin$.  \citet{SF10}
conclude that this dispersion profile clearly deviates  from the ``Newtonian prediction'', by which they
mean a monotonically declining dispersion profile, and is ``best explained by a breakdown of
Newtonian dynamics below a critical acceleration''.

Clearly there is an urgent need for additional accurate radial velocities for {\it bona-fide} members of
$\omega$~Cen in the outskirts of the cluster, particularly beyond $r$ $\approx$ 30$\arcmin$.
The generation of such a sample and a redetermination of the line-of-sight velocity dispersion 
profile in the outer regions of $\omega$~Cen is the purpose of this paper.  The sample 
selection, the observations, and the measurement of the radial velocities are discussed in the next section.
Section 3 discusses the membership status of the candidates, which is important given the low
density of $\omega$~Cen members in the outer regions of the cluster.  Section 4 presents the velocity 
dispersion profiles derived from both the new observations and from combining the new data with the 
velocities given by \citet{So09} and \citet{SF10}.
The results are presented and discussed in \S 5.  

\section{Observations and Reductions}

\subsection{Sample Selection}

The list of candidates to be observed consisted first of the 154 probable $\omega$ Cen members 
identified in DC08.  These stars lie between 20$\arcmin$ and 55$\arcmin$ from the cluster center,
possess velocities and line strengths consistent with cluster membership, and have 
15.4 $\leq$ $V$ $\leq$ 16.75 (DC08).  Then, in order to increase the sample of potential members in the 
outer parts of the cluster, two further lists were generated.  The first was simply the stars in the 
original DC08 sample that lie between 30$\arcmin$ and 60$\arcmin$ from 
the cluster center and which were not observed in DC08.  There are 545 candidates in this category.  
Note that the
outer radius limit of 60$\arcmin$ was chosen to match with the 2 degree diameter field-of-view
of the fiber-positioning system at the prime focus of the AAT, enabling fiber configurations to be
centered on the cluster.  It also matches the ``tidal radius''  of $\omega$ Cen.   

The second candidate list was generated from the same photometry set as used in DC08 but the selection 
window
was extended $\sim$0.5 mag fainter parallel to the cluster sequence in the color-magnitude diagram (see
Fig.\ 1 of DC08).  As for the brighter sample, the selected stars lie between 30$\arcmin$ and 60$\arcmin$ 
from the cluster center.  There are 1798 additional candidates is this list.

\subsection{Observations}

Five nights in 2008 April were allocated to this program with the 2dF fibre positioner and the AAOmega
spectrograph on the 3.9m Anglo-Australian Telescope (AAT)\@. 
The fibre positioner at the AAT prime focus can allocate a maximum of 392 fibers within the 2 degree 
diameter field-of-view.  Each fibre configuration then consisted of 6-8 guide fibre bundles, $\sim$300
fibres allocated to candidate members, up to 50 fibres allocated to blank-sky and $\sim$20 fibres
to likely members from DC08.  These latter stars were incorporated into every configuration to monitor for
any systematic effects in the radial velocities.  The fibres from the positioner are fed to the AAOmega
spectrograph -- a double beam instrument with separate blue and red cameras \citep{WS04, RS06}.
The spectrograph was configured with the $\lambda$5700\AA\/ dichroic and the 1500V (blue) and 
1700D gratings (red).  The blue spectra cover the wavelength interval 
$\lambda\lambda$4940--5650\AA\/ at a resolution $\lambda$/$\Delta\lambda$ of $\sim$4000.  
The red spectra were centred at $\lambda$8600\AA\/ with coverage from $\lambda$8340\AA\/ to  
$\lambda$8775\AA\/ including the Ca {\small II} triplet lines at $\lambda$8498, 8542 and 8662\AA.  
The resolution is $\sim$10,000 and the scale corresponds to 8.5 km s$^{-1}$ per pixel. In this paper 
we concentrate on the red camera data only. 

Less than ideal weather meant that only 11 $\omega$ Cen candidate member fibre configurations 
were observed, principally on 2008 April 25 and 2008 April 26.  Two of these configurations were 
repeats to compensate for diminished signal due to cirrus affecting earlier observations.    
Each configuration was observed as a set of 3 $\times$ 1500 sec exposures preceded by
fibre-flat and arc lamp exposures and followed by a second arc lamp exposure.  A number of bright radial 
velocity standards were also observed through individual fibres during the run.  These provide template
spectra for the cross-correlation analysis used to determine radial velocities.  Each dataset was reduced
with the pipeline reduction code 
{\it 2dfdr}\footnote{www.aao.gov.au/AAO/2df/aaomega/aaomega\_software.html
\#drcontrol}
which generates a wavelength-calibrated sky-subtracted spectrum for each object fibre.  The relative fibre 
transmissions were set from the data using the {\it SKYLINE(MED)} option which makes use
of the significant flux in each raw spectrum from the numerous bright night-sky emission lines in the 
wavelength region covered by the red camera data.  The individual spectra from each integration were 
then median combined to remove cosmic-ray contamination.

An additional configuration made up of likely new members as determined from the April observations 
was observed on 2008 May 31 via the AAT Service Observing program.  The instrumental set-up for
the service observations on the red side of AAOmega was the same as for the April observations and the 
data were observed and reduced in an identical fashion.  

\subsection{Radial Velocities}

Radial velocities were determined by cross-correlation techniques using the IRAF routine {\it fxcor}.
The template was a high signal-to-noise spectrum of the $V$ = 7.7 F6V star HD160043, which
provides a good match to the $\omega$ Cen member spectra, particularly as regards the width and
depth of the Ca {\small II} triplet lines.  The wavelength interval used for the correlation was 
$\lambda\lambda$8470--8740\AA\/ which encompasses the Ca {\small II} triplet lines as well as a
number of weaker lines, while minimising regions of potential significant residual from the sky subtraction.
After the cross-correlations were computed,  the output velocity
error and the cross-correlation peak height were plotted against the continuum count level in the 
correlation wavelength region for each of the 12 observed configurations.   
This enabled the identification of occasional situations where the
correlation had been affected by instrumental effects such as inadequate cosmic-ray removal.  In such
cases the problem was corrected, usually by interpolating over the effected pixels, and the
cross-correlation repeated.  In the final analysis velocities that had output errors exceeding 
5 km s$^{-1}$ and/or cross-correlation peak heights less than 0.5 were discarded -- these always 
coincided with the lowest signal-to-noise spectra.  

After applying appropriate heliocentric corrections, the zero point of the velocity system was determined
by correlating the template spectrum with other observations of the same standard in different fibres
and with similar spectra of three other radial velocity standards (HD83516, HD101266 and HD162356).
The zero point was set by minimising the difference between the observed relative velocities 
for these stars and their catalogue values.  The 11 observations of the four standards then have a 
standard deviation of 0.8 km s$^{-1}$ about their catalogue values indicating the velocity zero point is
well determined.  Only the 245 stars with velocities exceeding 100 km s$^{-1}$ were then retained for
the subsequent analysis.

To determine the velocity errors the rms deviation about the mean velocity was first calculated for the
$\sim$20 $\omega$ Cen members observed in the majority, if not all, of the 12 configurations.
Both the mean velocity and the rms were calculated using the output velocity errors from {\it fxcor} as
weights.  These data show that the rms about the mean velocity is below 1 km s$^{-1}$ when the 
continuum level in the cross-correlation region exceeds $\sim$900 ADU\@.  The rms values then
rise relatively rapidly with decreasing continuum levels to 1.7 $\pm$ 0.3 (1$\sigma$) km s$^{-1}$ 
at $\sim$600 ADU\@.
A similar analysis of the stars observed in the two repeated configurations is consistent with these values
and indicates further that the rms continues to rise with decreasing continuum level  to 
$\sim$3 $\pm$ 0.7 (1$\sigma$) km s$^{-1}$
at $\sim$200 ADU, the lowest continuum level of the stars remaining in the data set.  For stars with a 
single observation the velocity error was then set by this (rms, continuum level) relation while for stars
with multiple observations the error was taken as the rms divided by the square-root of the number of
observations.  The overall median velocity error is less than 1 km s$^{-1}$ excluding any systematic
zero point uncertainty.  This velocity error is lower than the 2--3 km s$^{-1}$
velocity error listed by \citet{RL09} who used a similar observing setup, although \citet{RL09} do not
give any information on the continuum levels of their spectra.  The uncertainty in the velocity errors
is sufficiently small that its contribution to the uncertainty in the calculated velocity dispersion is 
negligible. 

\section{$\omega$ Cen Membership}

\begin{figure}
\centering
\includegraphics[angle=-90.,width=0.9\textwidth]{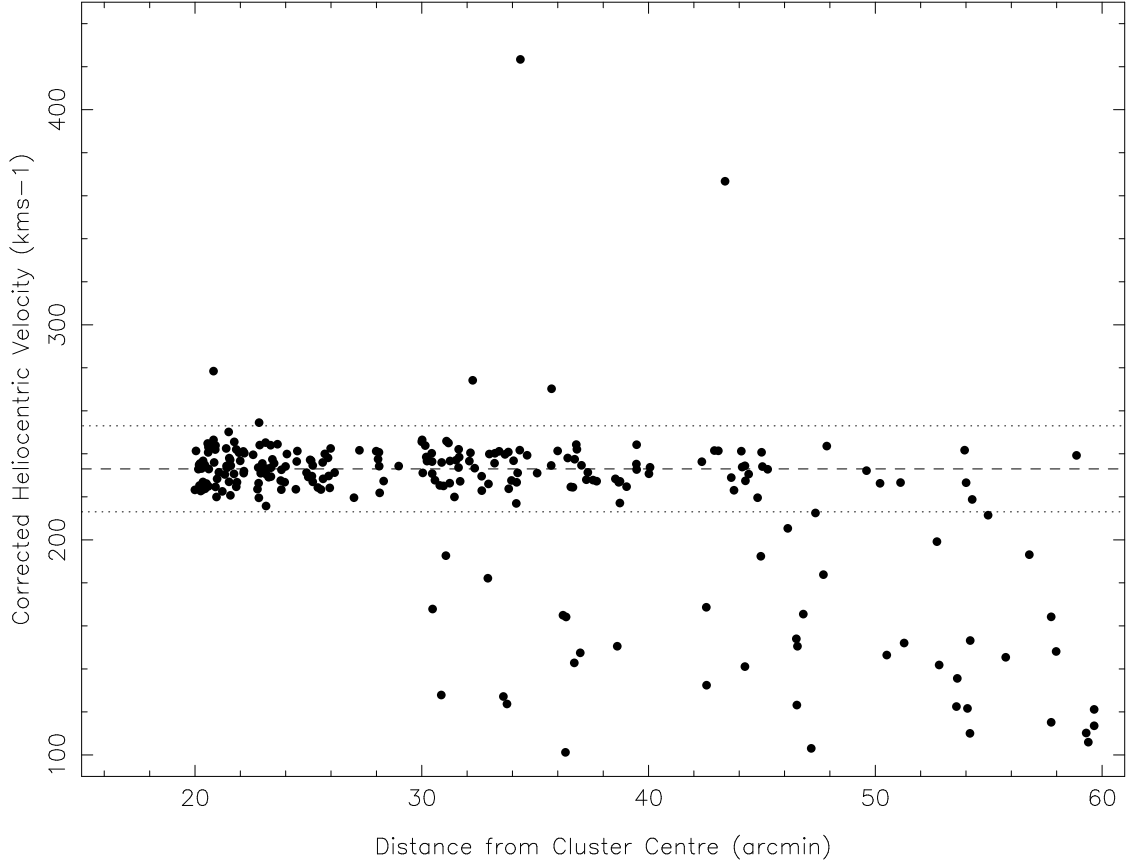}
\caption{The radial velocity, corrected for perspective rotation, is plotted against distance from
the center of $\omega$ Cen for all stars observed whose velocities exceed 100 km s$^{-1}$.  
Stars between
20$\arcmin$ and 30$\arcmin$ come from the sample of probable members of \citet{DC08} while
the stars at larger radii includes stars from that sample as well as stars observed here for the first time.  
The dashed
line is at the cluster mean velocity of 233 km s$^{-1}$ and the dotted lines are at $\pm$20 km s$^{-1}$
from the mean. \label{vels_all_fig}}
\end{figure}

In Fig.\ \ref{vels_all_fig} we plot the radial velocity, corrected for perspective rotation \citep[see][]{So09,
vdV06} against distance from the cluster center in arcmin.  The radial distances are computed 
using a tangent plane projection \citep[see][]{So09,vdV06} to allow for the large angular diameter of
the field surveyed.  
Considering first the 109 stars from DC08 with radial distances between 20$\arcmin$ and 30$\arcmin$,
it is evident that the vast majority of these stars are apparently probable cluster members despite the
relatively low velocity precision ($\sim$11 km s$^{-1}$) of the earlier study.  Only one star, 8\_7\_16862,
is definitely reclassified as a non-member based on the velocity determined here of  278.8 km s$^{-1}$,
while a second star, 8\_3\_1066 with v$_r$ = 254.2 km s$^{-1}$, lies just outside the $\pm$20 km s$^{-1}$
from the cluster mean boundaries shown in the figure.  This star is retained as a possible member for 
the moment.

For the 136 stars with v$_r$ $\geq$ 100 km s$^{-1}$ and which lie beyond 30$\arcmin$ from the 
cluster center, \mbox{Fig.\ \ref{vels_all_fig}} shows that there is an apparent grouping around 
the cluster mean velocity out
to a radius of at least 45$\arcmin$ and possibly beyond.  However, it must be kept in mind that the
surface density profile of the cluster is dropping rapidly with increasing radius: the 
profile given in DC08 indicates that the cluster star density drops by a factor of 5 between 20$\arcmin$
and 30$\arcmin$ and by a further factor of 10 between 30$\arcmin$ and 40$\arcmin$.  Conversely, the
area that needs to be surveyed goes up as $r^2$, as does the number of contaminating non-members
assuming they have a uniform surface density.  For these reasons coincidence with the $\omega$ Cen
mean velocity does not guarantee cluster membership in the outer regions of the cluster, and additional
information must be used to help exclude non-members.
Fortunately, we can make use of the known properties of the stellar population of $\omega$ Cen to
carry out this task.

The stellar system $\omega$ Cen is well known for its internal spread in [Fe/H] abundance.
\citet{JP10} have provided [Fe/H] values for a large sample of $\omega$ Cen red giants and their
results show, in agreement with earlier work \citep[see][for references]{JP10}, that: (a) there is a
lower bound at [Fe/H] $\approx$ --1.9 to the abundances of $\omega$ Cen red giants  that remains
constant with increasing radius; (b) there is a decrease in the number of stars with [Fe/H]
$\geq$ --1.3 with increasing radius relative to the number of more metal-poor stars; and (c) the
dispersion in abundance for stars with [Fe/H] $\leq$ --1.3 is approximately constant with radius.
The \cite{JP10} sample reaches out to radial distances of only $\sim$24$\arcmin$ but the limited
sample of more distant stars in \cite{JN97} suggest that these results apply also at larger radii.
Further, despite the complexity of the $\omega$
Cen color-magnitude diagram (CMD) in the vicinity of the main sequence turnoff \citep[e.g.][]{Be10}, it
is likely that the age-range among the cluster stars is comparatively small, less than 2--4 Gyr
\citep[e.g.,][and the references therein]{EP11}.  Thus there should be a reasonable level of 
consistency between overall
abundance inferred from location on the red giant branch in the CMD and that inferred
spectroscopically.  We can therefore use the photometry and the line strengths for the stars 
in the velocity window 213--253 km s$^{-1}$ to select probable members at all radial distances.  

\begin{figure}
\centering
\includegraphics[angle=-90.,width=0.55\textwidth]{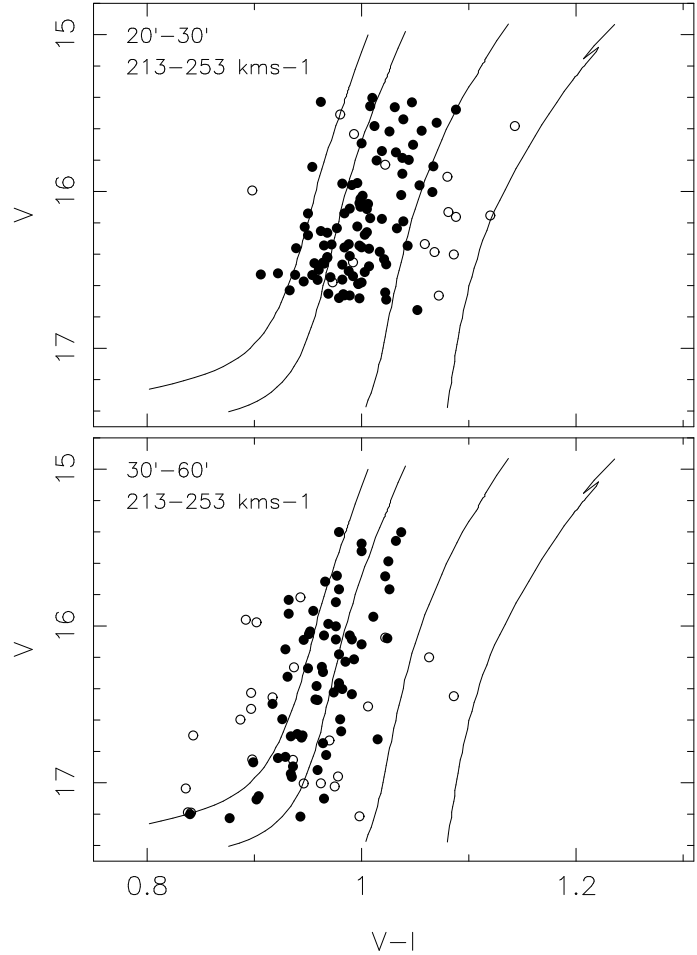}
\caption{Color-Magnitude Diagrams for stars with radial velocities between 213 and 253 km s$^{-1}$
and distances from the cluster center of 20$\arcmin$--30$\arcmin$ (upper panel) and 
30$\arcmin$--60$\arcmin$ (lower panel).  Shown also are Dartmouth isochrones for an age of 13 Gyr
and (left-to-right) ([Fe/H], [$\alpha$/Fe]) values of (--2.0, +0.4), (--1.5, +0.4), (--1.0, +0.4), and (--0.5, +0.2),
respectively.  The isochrones have been fitted assuming (m--M)$_{V}$ = 13.94 and E$(V-I)$ = 0.16 mag.
The stars adopted as cluster members are plotted as filled symbols while the likely non-members
are plotted as open symbols.
\label{members_fig1}}
\end{figure}

The approach is as follows.  In Fig.\ \ref{members_fig1} we show in the upper panel 
the CMD for the 108 stars in the velocity range 213--253 km s$^{-1}$  (plus 8\_3\_1066) which have 
radial distances between 20$\arcmin$ and 30$\arcmin$.   The lower panel shows the CMD for stars 
in the same velocity range but with radial distances between 30$\arcmin$ and 60$\arcmin$.  
These two groupings are designated the inner and outer samples,
respectively.  Note that for completeness we have included in the outer sample stars 5\_3\_226 and 
9\_4\_1918 as with velocities of 212.5 $\pm$ 1.2 and 211.4 $\pm$ 1.5 km s$^{-1}$, respectively, they
lie very close to the lower limit of the velocity selection range (see Fig.\  \ref{vels_all_fig}).
Shown also in both panels are theoretical
isochrones from the Dartmouth Stellar Evolution Database  \citep{Do08} for an age of 13 Gyr 
and metallicities [Fe/H]  of --2.0, --1.5,
--1.0 and --0.5 dex, respectively.  For the three lower metallicities the isochrones are for [$\alpha$/Fe]
= +0.4 while the most metal-rich isochrone has [$\alpha$/Fe] = +0.2 dex.  This is consistent with the
dependence of [$\alpha$/Fe] on [Fe/H] in the cluster \citep[e.g.,][]{EP02,JP10}.  The isochrones have been
fitted assuming (m--M)$_{V}$ = 13.94 and E$(V-I)$ = 0.16 mag (E$(B-V)$ = 0.12 mag) as tabulated
in the 2010 version of the Milky Way Globular Cluster database \citep[][hereafter H10]{WEH96}.

\begin{figure}
\centering
\includegraphics[angle=-90.,width=0.95\textwidth]{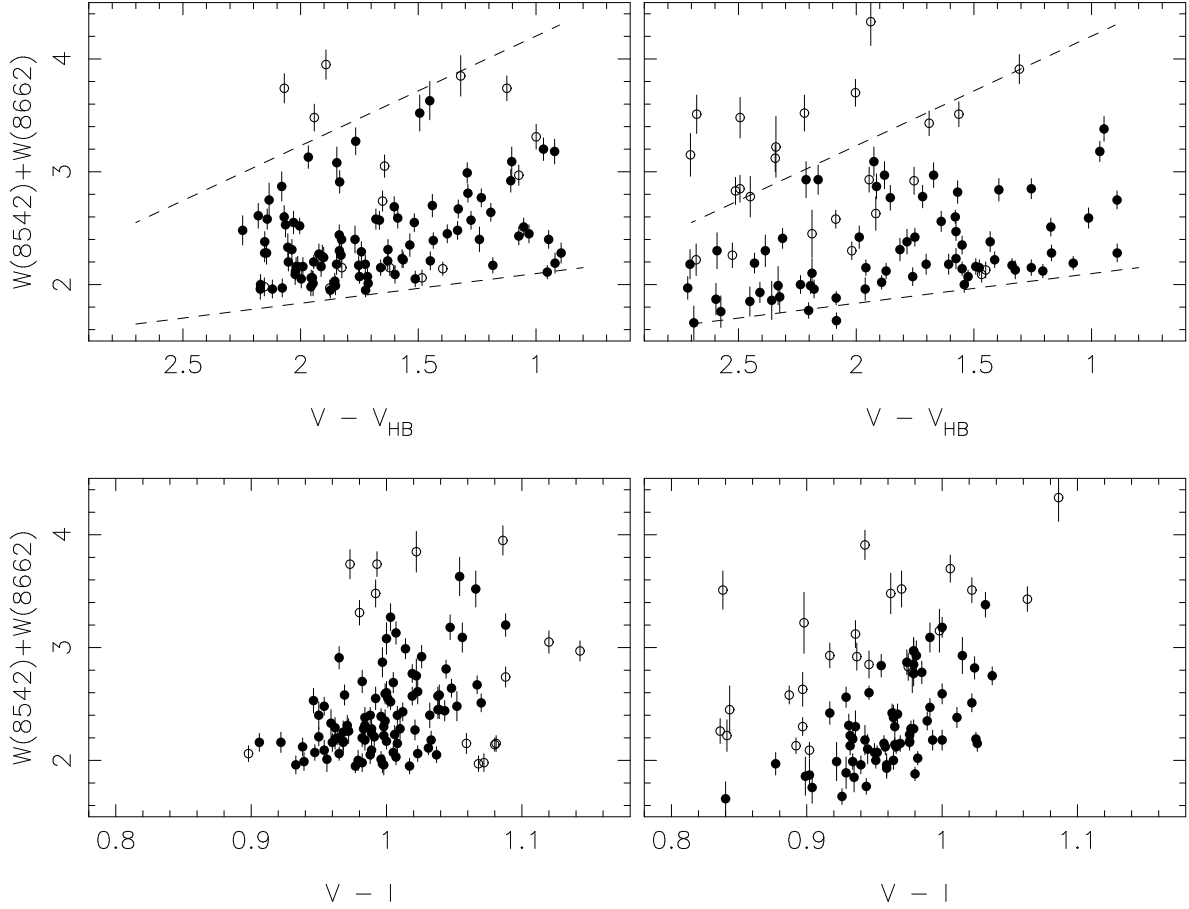}
\caption{The sum of the equivalent widths in \AA~of the Ca II triplet lines at
$\lambda$8542\AA~and $\lambda $8662\AA~lines is plotted
against $V-V_{HB}$ (upper panels) and $V-I$ (lower panels) for the inner sample (left panels)
and the outer sample (right panels).   The linear segments in the upper left
panel encompass the range of line strengths shown by probable cluster members in the inner sample.
The segments are replicated in the upper right panel.  The stars adopted as cluster members are plotted 
as filled symbols while the likely non-members are plotted as open symbols.
\label{members_fig2}}
\end{figure}

Similarly we show in Fig.\ \ref{members_fig2} plots of the combined equivalent widths of the 
$\lambda$8542\AA~and $\lambda$8662\AA~lines of the Ca II triplet against $V-V_{HB}$ and $V-I$ 
for the inner and outer samples.  The equivalent widths have been determined using gaussian fits to
the line profiles with feature and continuum bandpasses similar to those adopted in the original
work of \citet{AD91}.  The $V_{HB}$ value for $\omega$ Cen was taken from H10.  The values of
$V-V_{HB}$ lie outside the range of existing abundance calibrations for the Ca II triplet which are 
tailored to more luminous red giants.   Nevertheless we can use the inner sample, which is 
dominated by cluster members, to define reasonable upper and lower envelopes for the relation
between line strength and $V-V_{HB}$ followed by $\omega$ Cen stars.  The adopted linear relations 
are shown as the dashed lines in the upper left panel of Fig.\ \ref{members_fig2}.
The relations are then duplicated for the outer sample as shown in the upper right panel of the figure.
In both cases the lower envelope is very well defined consistent with the result for more luminous
samples that the lower abundance cutoff in the abundance distribution is quite sharp
\citep[e.g.,][]{JN96,JP10}.  The upper envelope is less well defined, as expected, given the significant
range of metallicities present even at large radial distances \citep{JN97}.  The lower panels
of Fig. \ref{members_fig2} show the relation between the Ca II triplet strength and $(V-I)$ colour for the
two samples.  Members of 
$\omega$ Cen should show a broad correlation between these quantities with, in general,
stronger lines going with redder colours.  The panels show this is the case.

\begin{figure}
\centering
\includegraphics[angle=-90.,width=0.55\textwidth]{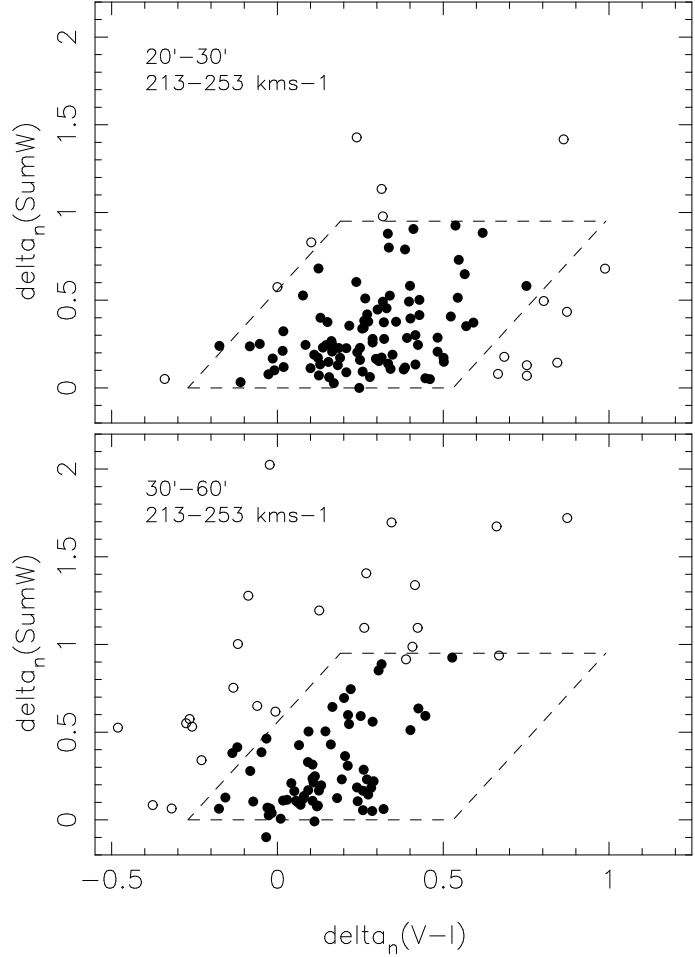}
\caption{The normalized relative line strength $\delta_{n}\Sigma$W in \AA~is plotted against
normalized relative giant branch color $\delta_{n} (V-I)$ for the inner sample (upper panel)
and for the outer sample (lower panel).  The dashed lines in the upper panel outline the 
membership selection criteria adopted for the inner sample.  The lines are reproduced in the lower 
panel.  The stars adopted as cluster members are plotted 
as filled symbols while the likely non-members are plotted as open symbols.  Stars near the
boundaries were considered individually and classified taking into consideration uncertainties
in the photometry and line strength measurements.
\label{members_fig3}}
\end{figure}

We then combine the information in Figs.\ \ref{members_fig1} and \ref{members_fig2} as
follows.  For each star in the panels of Fig.\ \ref{members_fig1} we
measure the offset between the $V-I$ color of the star and the color of the 13 Gyr, [Fe/H] = --2.0,
[$\alpha$/Fe] = +0.4 isochrone at the star's $V$ magnitude, normalized by the color difference
between this isochrone and that for 13 Gyr, [Fe/H] = --0.5, [$\alpha$/Fe] = +0.2 at the $V$ mag of the star.
For the same star we then measure the difference in equivalent width between the value for the star
and the lower envelope line shown in the upper panels of Fig.\ \ref{members_fig2} at the 
$V-V_{HB}$ value for the star.  This difference is then normalized by the equivalent width difference 
between the upper and lower lines at the $V-V_{HB}$ value of the star.  For members of the cluster these
two quantities, denoted by $\delta_{n}(\Sigma$W) and $\delta_{n} (V-I)$ respectively, should be well 
correlated as both are metallicity indicators.  Non-members, however, will in general lie away from the 
cluster member sequence.
The results of this process are shown in Fig.\ \ref{members_fig3} where the expected correlations are
evident.  The dashed lines in the upper panel of the figure show the membership selection 
window adopted for the inner sample.  The selection window is then reproduced in the lower panel
for the outer sample.

\begin{figure}
\centering
\includegraphics[angle=-90.,width=0.9\textwidth]{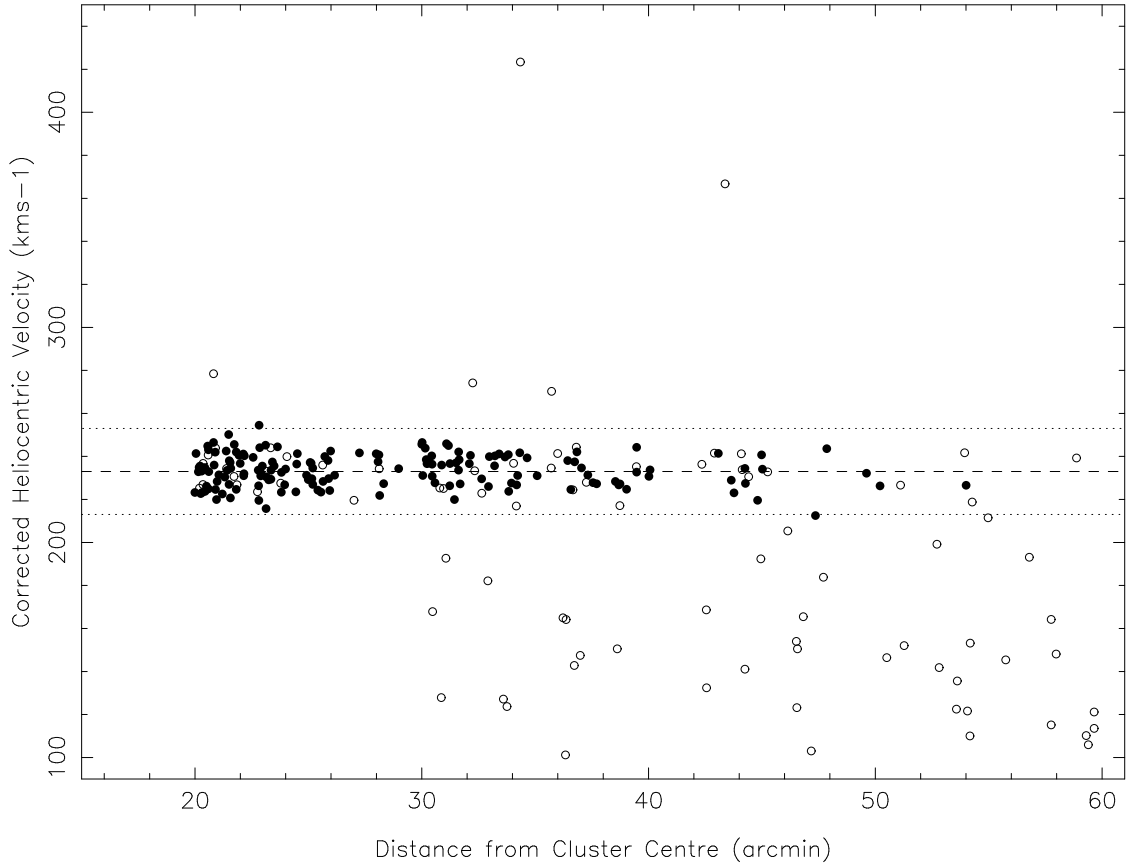}
\caption{As for Fig.\ \ref{vels_all_fig} the radial velocity, corrected for perspective rotation, is plotted 
against distance from the center of $\omega$ Cen for all stars observed whose velocities exceed 
100 km s$^{-1}$.   Adopted probable members are now plotted as filled symbols while probable 
non-members are plotted as open symbols. \label{vels_all_fig2}}
\end{figure}

We have then combined the information from Figs.\ \ref{members_fig1}, \ref{members_fig2} and
\ref{members_fig3} to provide our best estimate of the $\omega$ Cen membership status
for the stars in the inner and outer samples.  The adopted cluster members are plotted as
filled symbols in all three figures while the non-members are plotted as open symbols.  We note
that the membership status of stars falling near the boundaries of the selection window in 
Fig.\ \ref{members_fig3} were individually considered and classified taking into consideration 
uncertainties in the photometry and line strength measurements.  In particular, 
despite our efforts to minimise them it is still likely that there are
systematic uncertainties in the $(V-I)$ photometry at the $\pm$0.03 mag level.
For inner sample, 93 of the original 108 stars are classified as members, including star 8\_3\_1066,
while for the outer sample, 67 of the original 91 are classified as members, including 5\_3\_226
but not 9\_4\_1918.  
We then list  in Table \ref{Table1} the identification, J2000 position, heliocentric velocity and error, 
distance from the cluster center in arcmin, $V$ and $V-I$ photometry, and the sum of the equivalent 
widths of the $\lambda$8542\AA~and $\lambda $8662\AA~Ca II triplet lines in \AA, together with its 
associated error,  for the 160 adopted cluster members.  Table \ref{Table2} gives the same information
for 39 stars from the inner and outer samples that are classified as probable 
non-members of $\omega$~Cen.  Figure \ref{vels_all_fig2} shows the observed velocity versus
radial distance diagram of Fig.\ \ref{vels_all_fig} but now with the probable members and 
non-members identified.

\begin{deluxetable}{lccccccccc}
\tablewidth{0pt}
\tablecaption{$\omega$ Cen Probable Member Data \label{Table1}}
\tablecolumns{10}
\tablehead{
\colhead{ID} & \colhead{RA (2000)} & \colhead{Dec (2000)} & \colhead{V$_r$} 
& \colhead{$\sigma($V$_r$)}
 & \colhead{r$\arcmin$} & \colhead{$V$} & \colhead{$V-I$} & \colhead{$\Sigma$W} 
 &  \colhead{$\epsilon$} \\ 
& & & \colhead{(km s$^{-1}$)} & \colhead{(km s$^{-1}$)} & & \colhead{(mag)} & \colhead{(mag)}
& \colhead{(\AA)}  & \colhead{(\AA)}    }
\startdata
8\_7\_15831   & 13 25 06.54 & --47 17 41.6 &   223.4 & 0.4 & 20.0 & 15.80 & 1.01  & 2.99 & 0.09 \\ 
8\_4\_3206     & 13 27 47.33 & --47 45 44.2 &   241.6 & 1.0 & 20.1 & 16.42 & 0.97  & 2.25 & 0.10 \\
8\_8\_3134     & 13 25 47.75 & --47 10 59.8 &   232.6 & 0.7 & 20.2 & 16.69 & 1.02  & 2.61 & 0.11 \\
8\_6\_16385   & 13 24 56.26 & --47 36 34.7 &   235.7 & 0.6 & 20.2 & 16.38 & 1.02  & 1.95 & 0.07 \\
8\_2\_1336     & 13 28 43.30 & --47 24 37.8 &   234.8 & 0.4 & 20.2 & 16.19 & 1.04  & 2.58 & 0.09 \\
\enddata
\tablecomments{This table is available in its entirety in a machine-readable form in the online
journal.   A portion is shown here for guidance regarding its form and content.  For stars in common 
these data supercede those in Table 2 of Da Costa \& Coleman (2008).}
\end{deluxetable}

\begin{deluxetable}{lccccccccc}
\tablewidth{0pt}
\tablecaption{$\omega$ Cen Probable Non-Member Data \label{Table2}}
\tablecolumns{10}
\tablehead{
\colhead{ID} & \colhead{RA (2000)} & \colhead{Dec (2000)} & \colhead{V$_r$} 
& \colhead{$\sigma($V$_r$)}
 & \colhead{r$\arcmin$} & \colhead{$V$} & \colhead{$V-I$} & \colhead{$\Sigma$W} 
 &  \colhead{$\epsilon$} \\ 
& & & \colhead{(km s$^{-1}$)} & \colhead{(km s$^{-1}$)} & & \colhead{(mag)} & \colhead{(mag)}
& \colhead{(\AA)}  & \colhead{(\AA)}    }
\startdata
8\_8\_2219   & 13 26  04.35 & --47  09 39.7 &   224.8 & 0.7 & 20.2 & 16.40 & 1.09  & 3.95 & 0.13 \\
8\_8\_4146   & 13 25 24.37 & --47 13 35.6 &   226.8 & 0.8 & 20.4 & 16.15 & 1.12  & 3.05 & 0.10 \\
8\_1\_2780     & 13 27 51.11 & --47 11 28.5 &   235.7 & 0.3 & 20.4 & 15.83 & 1.02  & 3.85 & 0.18 \\
8\_8\_4052  & 13 25 26.75 & --47 12 55.7  & 240.5 & 0.4  & 20.6 & 16.13 & 1.08 &  2.15 & 0.07 \\
7\_3\_412     & 13 24 42.89 & --47 30 51.8 &   244.6 & 0.4 & 20.9 & 15.99 & 0.90  & 2.06 & 0.07 \\
\enddata
\tablecomments{This table is available in its entirety in a machine-readable form in the online
journal.   A portion is shown here for guidance regarding its form and content.  For stars in common 
these data supercede those in Table 2 of Da Costa \& Coleman (2008).}
\end{deluxetable}

We have then used the Besancon model of the Galaxy \citep{Ro03} to estimate the success of this 
membership selection process.  Five independent realisations of the model Galaxy were 
generated for the
line-of-sight towards $\omega$~Cen using an area on the sky equivalent to that of the 
2dF field-of-view.  The $V$ and $V-I$ magnitude and color ranges for the model
output were chosen to match 
approximately those of the outer sample of stars (see the lower panel of Fig.\ \ref{members_fig1}).
Normalisation of the model to the observational data was then set by the ratio of the number of stars in 
the outer sample with velocities in the range 140--190 km s$^{-1}$ to the number of model stars in the
same velocity interval.  The predicted number of field stars in the velocity interval 213--253 km s$^{-1}$
could then be calculated from the model numbers.  The predicted number of field stars with 
velocities exceeding 260 km s$^{-1}$ was also calculated as a check.  We find that the model
normalised in this way over predicts the number of high velocity stars: 8$\pm$1 stars are predicted
versus the 4 actually observed in the outer sample.  The difference however, is not very
significant given the small numbers.

For the cluster velocity range, the normalised model predictions are 3$\pm$1 field stars for distances 
from the cluster center between 30$\arcmin$ and 40$\arcmin$, 2$\pm$1 field stars for between 
40$\arcmin$ and 50$\arcmin$, and 3$\pm$1 field stars for between 50$\arcmin$ and 1$\arcdeg$.  
The number of stars classified as probable non-members in the corresponding radial distance intervals
(see Fig.\ \ref{vels_all_fig2}) are 13, 6 and 4 stars, respectively.  Assuming the validity of the model 
and the normalisation, the comparison then appears to
indicate that the membership classification process adopted has in all likelihood been too
conservative: perhaps as many as dozen of the ``probable non-members'' are in fact likely to be
cluster members, with most of the mis-classified stars falling in the 30$\arcmin$--40$\arcmin$
radial range.  The effect of this potentially overly conservative membership selection on the calculated
velocity dispersions will be discussed in the following section. 

In Fig.\ \ref{vels_all_fig} there are eight stars with velocities between 213 and 253 km s$^{-1}$, 
plus stars 5\_3\_226 and 9\_4\_1918 which lie just outside the velocity interval, that have
distances from the cluster center exceeding 46$\arcmin$.  Only five of these ten stars survive the cluster
membership analysis and are listed in Table \ref{Table1} and shown as filled symbols in 
Fig.\ \ref{vels_all_fig2}.  This number is too small for a statistically
meaningful measure of the velocity dispersion at this extreme outer region and so the subsequent 
analysis will be based on the 155 probable $\omega$~Cen members that lie between 20$\arcmin$ 
and 46$\arcmin$ from the cluster center.

\begin{figure}
\centering
\includegraphics[angle=-90.,width=0.6\textwidth]{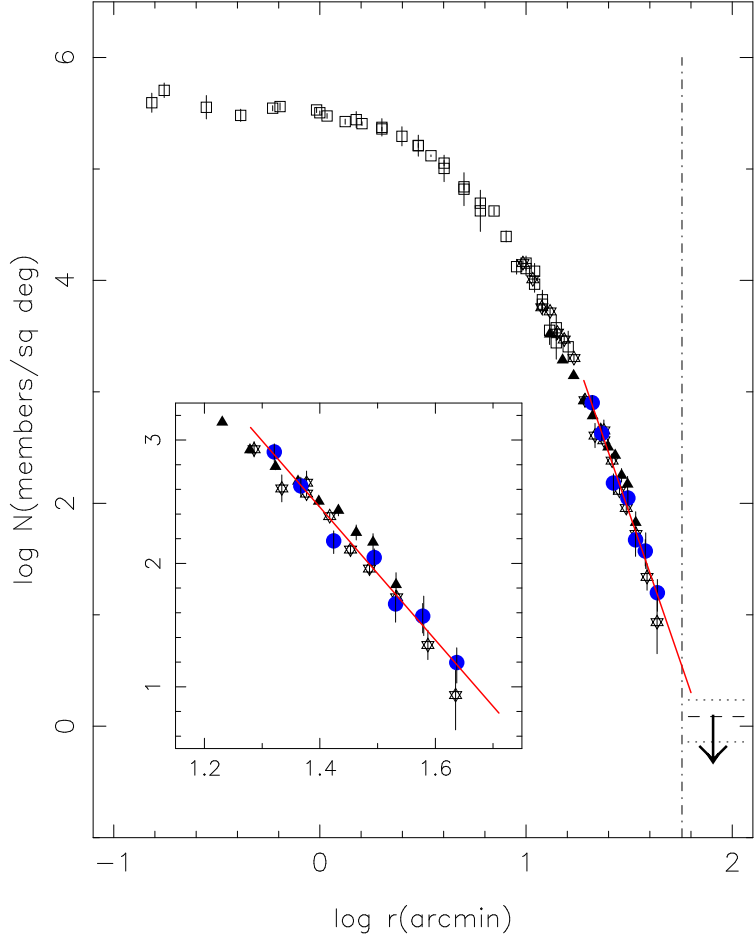}
\caption{The (circularly averaged) surface density profile for $\omega$~Cen taken from
\citet{DC08} except that the blue filled circles, which represent the cluster member sample
determined here, replace the equivalent data in the earlier work.  The vertical dot-dash line indicates
the tidal radius adopted by \citet{DC08}.  The dashed line and the vertical arrow indicate the upper
limit on the density of cluster members for the region between 1 and 2 tidal radii derived by \citet{DC08},
while the dotted lines represent the statistical uncertainty in the upper limit.  The red solid line is a
power law fit to the outer points; it has a slope of --5.4 dex/dex.  The fit to the outer points is enlarged
in the insert box. 
\label{paper_fig4}}
\end{figure}

Before discussing the velocity dispersion profile defined by these data we show in Fig.\ \ref{paper_fig4} 
the surface density profile for $\omega$~Cen.  The data are
taken from DC08 (see references therein) except that we have used the surface density values
determined from the cluster member sample derived here in place of the equivalent data in DC08.
The surface density points from the inner and outer samples have been separately scaled vertically
to match the existing data.  The separate scaling is necessary since the outer sample covers a
larger $V$ magnitude range than the inner sample.  
It is worth noting that the outer parts of this profile do not obviously
exhibit the increased profile steepening at large radii characteristic of a tidally limited profile.  Instead it
appears that the outer profile is best described by a constant slope, i.e., a power-law profile.  
A (unweighted) least-squares fit to the data points lying beyond 20$\arcmin$ (1.3 in log $r$) yields
a value of --5.4 $\pm$ 0.2 for the power-law slope.  The fit is shown as the straight line in the figure.
It is also worth noting that the surface density implied by the 5 possible members beyond 45$\arcmin$,
whose radial distances range between 47.4$\arcmin$ and 54.0$\arcmin$, is  consistent with
the observed profile and the power-law fit.   

\section{The Velocity Dispersion Profile}

The outer sample of \citet{So09} contains 98 $\omega$ Cen members that have distances from the
cluster centre exceeding $\sim$20$\arcmin$, although there are only 13 members beyond 30$\arcmin$.
Similarly, \citet[][see also \citet{SM03}]{SF10} give radial velocities for 75 $\omega$ Cen members with 
radial distances beyond 20$\arcmin$ from the cluster center, but there is only one 
member (just) beyond 30$\arcmin$ in that sample.  There are 26 stars in common between the 
\citet{So09} and 
\citet{SF10} samples.  The current sample, with 62 probable members between 30$\arcmin$ 
and 46$\arcmin$ therefore represents a considerable increase in the number of velocities with which 
to study the velocity dispersion profile in the extreme outer parts of this stellar system.

\subsection{AAT sample}

\begin{figure}
\centering
\includegraphics[angle=-90.,width=0.9\textwidth]{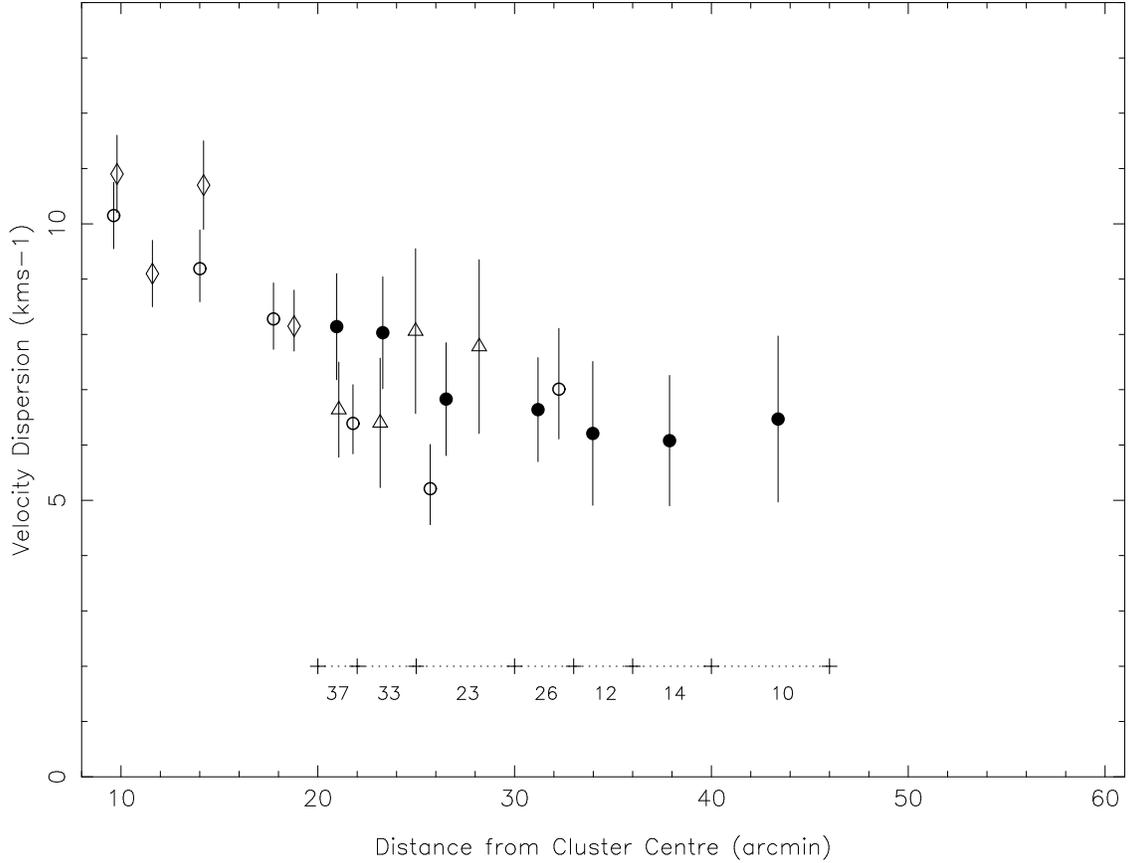}
\caption{The line-of-sight velocity dispersion of $\omega$ Cen member stars is plotted against
distance from the cluster center in arc minutes.  The filled symbols are for the current probable
members data set.
The radial range corresponding to each dispersion point is indicated in the lower part of the
plot as is the number of stars in each radius bin.  Shown also on the plot are the line-of-sight
velocity dispersion points of \citet{vdV06} (open diamonds), \citet{SM03} 
\citep[from][open triangles]{SF10},
and \citet{So09} (open circles).  There is no evidence for any significant decline in the velocity
dispersion beyond $\sim$25$\arcmin$.
\label{paper_fig5}}
\end{figure}

In Fig.\ \ref{paper_fig5} we show the velocity dispersion points calculated from the present sample
of $\omega$ Cen probable members.  
The velocity dispersions have been calculated using a maximum likelihood estimator 
\citep[e.g.,][]{PM93}.   In making the calculation
the stars have been grouped into bins containing at least 10 members -- the radial range of each bin
and the number of stars included are indicated in the lower part of the figure.  The dispersions have
been calculated relative to the mean velocity for each group -- these individual mean velocities differ by 
at most 2.3 km~s$^{-1}$ from the mean (233.4 km~s$^{-1}$) for the 93 members  with distances 
between 20$\arcmin$ and 30$\arcmin$ from the cluster center.  Calculating the dispersions for the
stars beyond 30$\arcmin$ relative to this fixed mean causes only a slight {\it increase} (0.1 to 0.4 
km~s$^{-1}$) in the dispersions but the differences are well within the uncertainties.

We also note that in the previous section it was suggested that the membership
selection employed had been too conservative in that perhaps as many as a dozen of the ``probable
non-members'' may be actual cluster members.  To investigate the effect of this possibility on the
calculated dispersions, we conducted five trials in which 12 stars were randomly selected from the
set of 19 probable non-members that lie in the radial range 30$\arcmin$  to 46$\arcmin$.  The
selected stars were then combined with the probable members in the appropriate radial bins and the 
dispersions recalculated.  In all cases the change in the dispersion was less than the errors 
calculated for the probable members only samples.  The mean change was an increase in
dispersion of 0.3 km~s$^{-1}$ with the largest excursions seen being an increase in the dispersion
in the 33$\arcmin$--36$\arcmin$ bin of 1.2 km~s$^{-1}$ and a decrease of 0.6 km~s$^{-1}$ in
the dispersion for the 40$\arcmin$--46$\arcmin$ bin.  Both these changes are within the error for
the equivalent probable members only sample.  We conclude therefore that the dispersion
measurements are stable against modest changes in the membership status of individual stars.

Shown also in Fig.\ \ref{paper_fig5} are the velocity dispersion measurements from 
\citet{So09}, taken directly
from their Table 1, from \citet{vdV06} as the line-of-sight dispersions from their Figure 8, and from 
\citet{SF10}.  In the latter case the dispersion points were calculated directly from the heliocentric
radial velocities listed in Table 2 of \citet{SF10} using identical techniques, including correction for
perspective rotation, as those employed for the $\omega$ Cen stars observed here.

It is apparent from Fig.\ \ref{paper_fig5} that within the radial range where the different samples 
overlap, the velocity dispersion measures are consistent with one another.  It is also evident that there
is no evidence for any significant decline in the line-of-sight velocity dispersion of $\omega$ Cen
members beyond $\sim$25$\arcmin$.  In particular, the new data, which extend well beyond
the previous data, are consistent with a constant line-of-sight velocity dispersion of $\sim$6.5 
km~s$^{-1}$ in the outer parts of the cluster.
 
\subsection{Combined Sample}

The 26 stars in common between the \citet{So09} and the \citet{SF10} samples have a mean 
velocity difference (\citet{SF10} -- \citet{So09}) of 0.2 km~s$^{-1}$, with a standard deviation of
0.8 km s$^{-1}$, confirming the velocity precision of both data sets.  Unfortunately, due to the 
difference in the apparent magnitudes of the samples, there are no stars in common between the
present work and that of \citet{So09}.  There are, however, three stars in the present sample
that are also in \citet{SF10}.  \citet{SF10} stars 0006, 78004 and 85007 correspond to stars 8\_8\_4776,
8\_5\_6453 and 7\_4\_160 here.  For the first two stars the velocity differences 
(\citet{SF10} -- present work) are gratifyingly small: --0.6 and --0.2 km~s$^{-1}$, respectively,
but for the third star
the \citet{SF10} velocity exceeds that found here by 11.7 km~s$^{-1}$.  Star 7\_4\_160 is one of the
stars observed in all 12 configurations here and there is no indication of any velocity variability: the rms
about the weighted mean velocity is 1.2 km~s$^{-1}$.  Consequently, we have not used the \citet{SF10} 
velocity for this star.  Otherwise, the velocities for the stars in common have been averaged.
The combined data set then has a total of 299 stars, 224 with radial distances between 20$\arcmin$
and 30$\arcmin$ and 75 lying between 30$\arcmin$ and 46$\arcmin$.

\begin{figure}
\centering
\includegraphics[angle=-90.,width=0.9\textwidth]{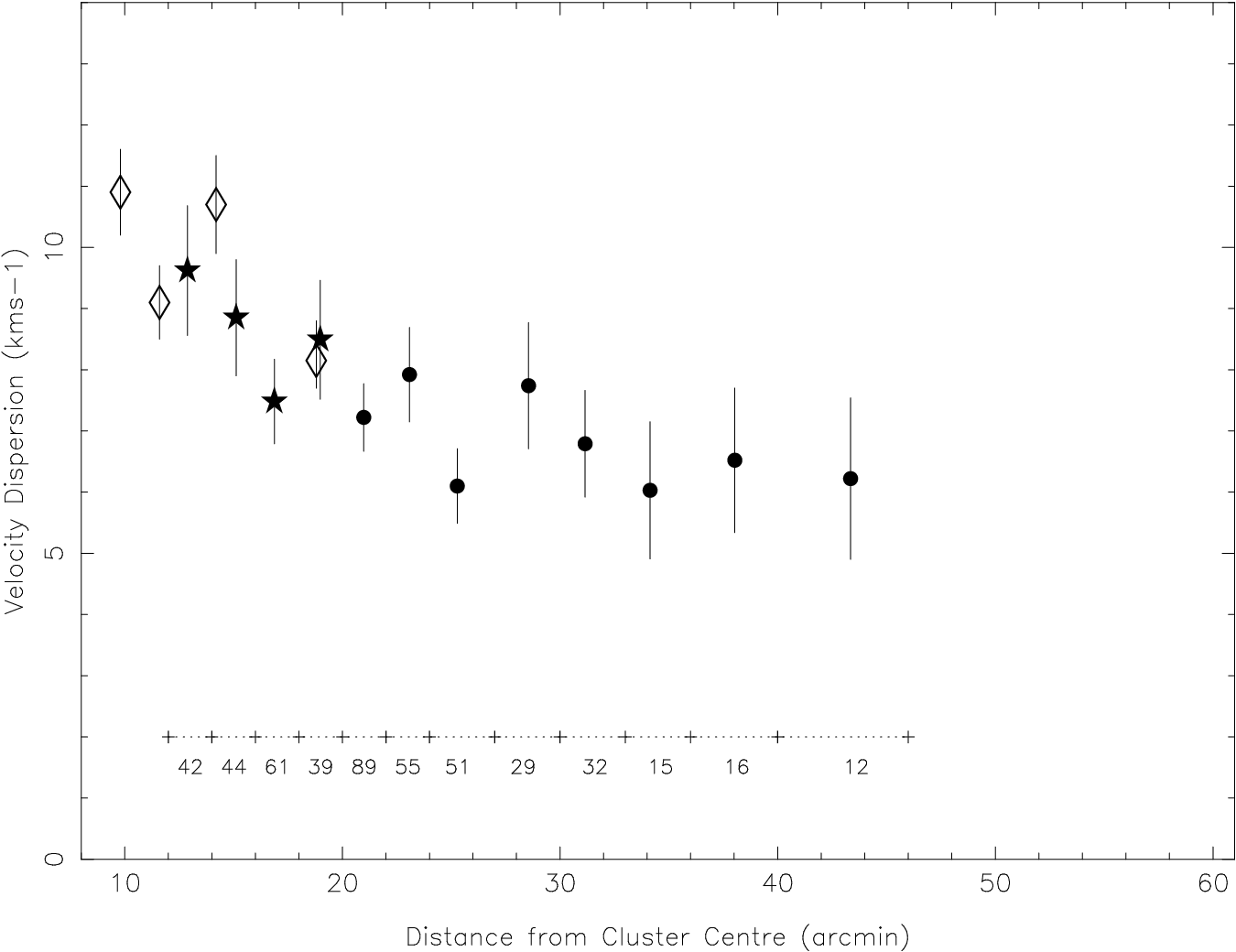}
\caption{The line-of-sight velocity dispersion of $\omega$ Cen member stars is plotted against
distance from the cluster center in arc minutes.  The filled circles are for the combined data
set of this work, \citet{So09}, and \citet{SF10}.  The filled stars are from the data set of 
\citet{So09} for stars between 12$\arcmin$ and 20$\arcmin$.  Line-of-sight
velocity dispersion points from \citet{vdV06} are shown as open diamonds.
For the combined and \citet{So09} points the radial range corresponding to each dispersion point 
is indicated in the lower part of the plot, as is the number of stars in each radius bin.  
As in Fig.\ \ref{paper_fig5}, there is no evidence for any significant decline in the velocity
dispersion beyond $\sim$25$\arcmin$.
\label{paper_fig6}}
\end{figure}

Figure \ref{paper_fig6} then shows the velocity dispersion profile for the combined sample.
Shown also in the figure are velocity dispersion measurements calculated from the sample of
\citet{So09} for stars between 12$\arcmin$ and 20$\arcmin$.  The data are tabulated in
Table \ref{Table3}.  The line-of-sight velocity dispersion
points from \citet{vdV06} shown in Fig.\ \ref{paper_fig5} are also reproduced in the figure.
The combined sample clearly verifies what was already evident from Fig.\  \ref{paper_fig5} -- that although
the velocity dispersion decreases outwards with increasing radius for the inner parts of the cluster
(see, for example, figure 8 of \citet{So09} or figure 4 of \citet{SF10}), beyond $\sim$25$\arcmin$ the
velocity dispersion profile shows no signs of decreasing with increasing radius.  This is despite, as
noted above, the surface density dropping by a factor of $\sim$10 between radial distances of
30$\arcmin$ and 40$\arcmin$.  For the 140 stars in the combined sample with $r$ $\geq$ 25$\arcmin$,
the mean radius is 31.3$\arcmin$ and the velocity dispersion is 6.6 $\pm$ 0.4 km~s$^{-1}$.

\begin{deluxetable}{ccccc}
\tablewidth{0pt}
\tablecaption{Velocity dispersion profile for the outer parts of $\omega$ Cen from the combined
sample \label{Table3}}
\tablecolumns{5}
\tablehead{
\colhead{Radius Range} & \colhead{Mean Radius} & \colhead{N} & \colhead{$\sigma$} 
& \colhead{error}\\
\colhead{(arcmin)} & \colhead{(arcmin)} & & \colhead{(km s$^{-1}$)} & \colhead{(km s$^{-1}$)} 
    }
\startdata
12--14 & 12.88 & 42 & 9.62 & 1.06 \\
14--16 & 15.12 & 44 & 8.85 & 0.95 \\
16--18 & 16.87 & 61 & 7.48 & 0.69 \\
18--20 & 18.98 & 39 & 8.49 & 0.97 \\
20--22 & 20.98 & 89 & 7.22 & 0.55 \\
22--24 & 23.08 & 55 & 7.92 & 0.77 \\
24--27 & 25.28 & 51 & 6.10 & 0.61 \\
27--30 & 28.55 & 29 & 7.74 & 1.03 \\
30--33 & 31.15 & 32 & 6.79 & 0.87 \\
33--36 & 34.14 & 15 & 6.03 & 1.12 \\
36--40 & 38.03 & 16 & 6.52 & 1.18 \\
40--46 & 43.35 & 12 & 6.22 & 1.32 \\

\enddata
\tablecomments{The first four entries are based on the data set of \citet{So09} while the remaining
entries are drawn from the combined data set of this work, \citet{So09} and \citet{SF10}.}
\end{deluxetable}

\section{Discussion}

The first question to be addressed is whether there is a dynamical model  which 
can reproduce the surface brightness/surface density profile of the cluster, and the observed
line-of-sight velocity dispersion
profile, {\it without} any requirement for dark matter (i.e., a model in which mass follows light) or 
similarly, without any requirement for
non-Newtonian gravity.  For example, the \citet{CW75}-type model for $\omega$~Cen presented in
\citet{So09} fits the projected surface density profile, the ellipticity profile and the rotation curve
adequately, as shown in Figure 9 of \citet{So09}.  However, while the model also reproduces the 
\cite{So09} velocity dispersion data, it does not fit the more extensive velocity dispersion profile 
data presented here.
The velocity dispersion profile of the model is consistently below the observed points in Fig.\
\ref{paper_fig6} beyond $\sim$25$\arcmin$, and declines monotonically to, for example, 3.6 km~s$^{-1}$
at $\sim$45$\arcmin$, significantly below the observations.  The model is therefore not an adequate
description of the dynamics in the outer parts of the cluster.  

This is likely to be the case for all similar models, e.g., \citet{MM95,MM03,MvM05}\footnote{The latter
two references show a observed velocity dispersion point of 3.5 $\pm$ 1.5 km~s$^{-1}$ at a radius
of 36.6$\arcmin$, which is in accord with the predictions of the Wilson-type models.  This datum comes
from \citet{PS83} and is based on velocities for four stars whose radial distances extend
between $\sim$31$\arcmin$ and $\sim$44$\arcmin$.  Three of the four stars are included in the sample 
of \citet{So09}.  We assert that the considerably more extensive data set presented here supercedes this
early result.}, primarily because 
they are based on a fundamental assumption
that the velocity distribution function $f(v)$ is of a `lowered Maxwellian' form \citep{K66,CW75}.  This
ensures that the density reaches zero at a finite radius, usually identified with the boundary set
by the tidal force of the Milky Way \citep{K66}.  In such models the velocity dispersion profile also
declines monotonically, reaching zero at the same finite radius \citep[e.g., Fig.\ 1 of][]{MM95}.  For 
$\omega$~Cen, however, the lack of an obvious tidal radius cutoff signature in the surface density 
profile (see Fig.\ \ref{paper_fig4})  suggests that models of this type are not appropriate 
for the outer parts of the stellar system.  In this sense the disagreement with the velocity dispersion
observations reveals the inadequacy 
of the models, not necessarily anything more fundamental.

A more heuristic approach to the modelling of $\omega$~Cen is that taken by \citet{vMA10}, whose
primary aim was to place constraints on the possible presence of an intermediate-mass 
black hole at the center of the cluster.  Their approach was to parameterise the surface density profile
and then solve the spherical anisotropic Jeans equation to predict the velocity dispersion profiles. 
The surface brightness profile was fit with a so-called
``generalised nuker'' profile \citep{vMA10}, which allows for a central power-law cusp and which contains
two characteristic logarithmic slopes with associated ``break'' radii \citep[see][for details]{vMA10}.  This
adopted functional form does not have any particular physical significance, but in this context 
it is important to note that a density profile of this type, in contrast to the \citet{K66} and 
\citet{CW75}-type models
described above, does not have a finite cutoff radius at which the density, and the velocity
dispersion, go to zero.

\begin{figure}
\centering
\includegraphics[angle=-90.,width=0.9\textwidth]{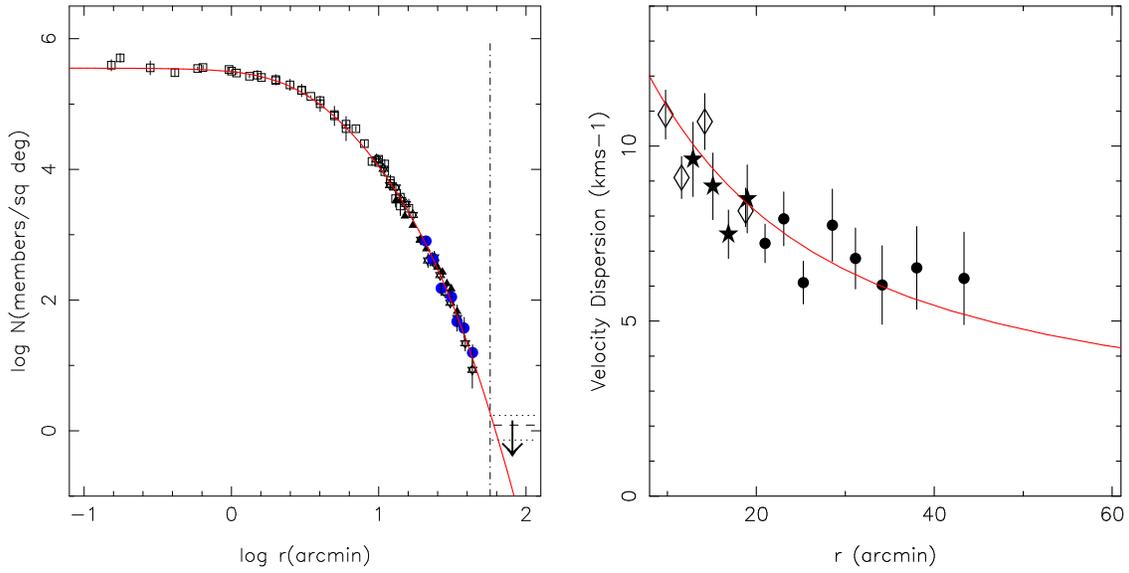}
\caption{{\it Left panel:} The surface density profile for $\omega$~Cen, from Fig.\ \ref{paper_fig4}.  
The red solid line is the surface density profile of the best-fit  ``generalised nuker'' model of \citet{vMA10} 
for the case with the central logarithmic slope $\gamma$ fixed at zero.  The model profile
has been scaled vertically to fit the surface density observations. {\it Right panel:} The line-of-sight 
velocity dispersion of $\omega$ Cen for the outer parts of the cluster.  Symbols are as for 
Fig.\ \ref{paper_fig6}.
The red solid line is the line-of-sight velocity dispersion profile of the same ``generalised nuker'' 
model whose surface brightness profile is shown in the left panel. 
\label{paper_fig7}}
\end{figure}

We show in the left panel of Fig.\ \ref{paper_fig7} the best-fit ``generalised nuker'' 
model of \citet{vMA10} for the case where 
the central logarithmic slope $\gamma$ is fixed at zero\footnote{This model differs from the overall 
best-fit model, which
has a shallow cusp with $\gamma$ = 0.05, only in the very inner regions of the cluster 
($r$ $\lesssim$ 20$\arcsec$).}, compared to the surface density data 
presented in Fig.\ \ref{paper_fig4}.  The model data have been scaled vertically to correspond to the
adopted surface density scale of the observations. 
The fit is excellent at all radii: in particular, the relatively constant logarithmic slope at large radii is
in agreement with the observations, noting that the new surface density data presented here, and those
given in DC08, were not included in the \citet{vMA10} fitting process.  The right panel shows the 
corresponding line-of-sight velocity dispersion profile compared with the observations from 
Fig.\ \ref{paper_fig6}.  The velocity distribution is mildly anisotropic with the transition from
radial anisotropy at small radii to tangential anisotropy at large radii occurring at $r$ $\approx$ 
12$\arcmin$ \citep{vMA10} in agreement with the results of \citet{vdV06}.  The model
velocity dispersion  curve is 
consistent with the observations over the entire radial range depicted, including the points beyond 
$\sim$25$\arcmin$.  For completeness we note that the
upper panel of Figure 7 of \citet{vMA10} shows that this model also 
reproduces satisfactorily the observed velocity dispersion profile in the inner parts of the cluster.
The model explicitly assumes `mass follows light' and yields a $V$-band mass-to-light ratio of 2.6
in solar units, in agreement with the M/L$_{V}$ value found in the \citet{vdV06} study.  These dynamical
values are entirely
consistent with the mass-to-light ratio expected, given plausible assumptions about the
stellar population of the cluster \citep[e.g.,][]{GM87}.  We can conclude therefore, based on
this parameterized model, that an extended dark matter distribution is not required to reproduce the
observed velocity dispersion profile in the outer parts of the cluster.  Similarly, given that the modelling
process is based on standard Newtonian dynamics through the use of the Jeans equation, 
no non-standard dynamics are required. 

The \citet{vMA10} model may well be a reasonable description of the dynamics in the outer parts of 
$\omega$~Cen but because of its heuristic nature it does not provide any direct insight into the 
physical processes responsible for the applicability of the model.  To investigate this we note again that 
the outer surface density profile of $\omega$~Cen is well represented by a power-law and does not
exhibit the characteristic ``King-profile'' tidal cutoff seen in many clusters \citep[e.g.,][]{MvM05,JG10}.  As 
discussed by \citet[][see also \citet{KK10}]{PN09} this is an indication that the phase space in the 
outer parts of the
cluster is likely populated up to $E$ $\sim$ 0, requiring a source of additional energy.
The likely source of the required heating is the tidal shocks that occur each time $\omega$~Cen 
crosses the Galactic plane, as well as the tidal heating the cluster experiences as it moves in the spatially
varying potential of the Galaxy.

\citet{Di99} used the current position and motion of $\omega$~Cen to characterise the orbit of the
cluster around the Galactic Center. They found that the system has peri- and apo-Galactocentric
distances of approximately 1.2 and 6.2 kpc, and an orbital period of $\sim$120 Myr \citep{Di99}.
Using these orbital parameters \citet{vdV06} calculate that the velocity component perpendicular to
the Galactic plane $v_\bot$ is of order 40 km~s$^{-1}$.  Consequently, for a disk scale height of 
$\sim$250 pc,
it takes $\omega$~Cen about 12 Myr to cross through the disk of the Galaxy \citep{vdV06}.
In contrast, in the $\omega$~Cen model of \citet{vdV06} the orbital timescale for member stars
in the outer parts of the cluster is approximately 100 Myr at $r$ $\sim$ 25$\arcmin$ -- 30$\arcmin$, and
longer at larger radii.  Thus the impulse approximation \citep[e.g.,][p.\ 446]{GLO99,BT87} is valid for
calculating the ``shock heating''  the outer parts of the cluster experience each disk crossing.

We use equation 7-71 of \citet{BT87}, with parameter values from \citet{vdV06}, to show that the
impulsive change in the velocities of stars, $|\Delta v|$, is $\sim$ 0.36 $r\arcmin$ km~s$^{-1}$.
This change is then comparable to, or exceeds, the line-of-sight velocity dispersion for
radial distances beyond $r$ $\approx$ 20$\arcmin$.
The relative importance of disk shock heating is then measured by a comparison of the shock heating
timescale $t_{shock}$ with the dynamical timescale $t_{dyn}$.  Use equation 7-72 of \cite{BT87} with
the parameter values from \citet{vdV06} gives:
\begin{equation}
t_{shock} = 475\sigma^2 / r^{2}
\end{equation}
for $t_{shock}$ in Myr, $\sigma$ in km~s$^{-1}$ and $r$ in arcmin\footnote{The value of the numerical
coefficient in this equation given by \citet{vdV06}, 21, is incorrect (G. van de Venn, {\it priv.\ comm.} 2012).
However, the error does not significantly affect the discussion in \citet{vdV06}, although the influence
of tidal shocks at a given radius are over-estimated in that paper.}.
Adopting $\sigma$ = 
6.5 km~s$^{-1}$ for $r$ $\gtrsim$ 25$\arcmin$ and then comparing the radial variation of
$t_{shock}$ with that for $t_{dyn}$ from Figure 21 of \citet{vdV06} shows that $t_{shock}$
$\sim$ $t_{dyn}$ at $r$ $\sim$ 27$\arcmin$ and that $t_{shock}$ $<$  $t_{dyn}$ at larger radii.   
Consequently, we can conclude that beyond $r$ $\sim$ 25--30$\arcmin$ 
the energy input to the outer parts of
cluster from the disk shocking process is significant, and it will increasingly dominate the dynamics
as the radius increases.  Indeed when $t_{shock}$ is less than the stellar orbital timescale  
\citep[$\sim$4$t_{dyn}$,][]{vdV06} the stars are unlikely to be in equilibrium with the cluster potential.   
It is also worth noting that at approximately the same cluster radius as where shock heating becomes
important, the stellar orbital timescale exceeds the orbital period of the cluster around 
the center of the Galaxy.  As a result, the outer parts of the cluster will also experience tidal heating 
due to the changing potential field of the Galaxy as the cluster moves from its apo- to peri-Galactic 
distances  \citep[e.g.,][]{GO97}.  The significance of these two effects
then suggests strongly that the phase space structure of the outer parts of $\omega$~Cen is 
dominated entirely by {\it external} effects driven by the cluster's location relatively close to the 
center of the Galaxy. The same conclusion was reached by \citet{vdV06}.

What is needed to shed further light on the situation is a full numerical simulation in
which the dynamics of an $\omega$~Cen-like system are explored as the system orbits in the 
potential of the Galaxy.  Such a calculation needs to include the effects of disk-shocking and continue
for a sufficient time that the quasi-equilibrium situation that likely applies in the outskirts of 
$\omega$~Cen becomes established.   The results of \citet{KK10}, for example, are suggestive in
this respect.   Based on $N$-body calculations of model star clusters with masses of a few 
10$^4$ M$_\sun$ on various orbits, \citet{KK10} demonstrate that tidal heating can lead 
to a population of ``potential escapers'', i.e., energetically unbound stars inside the cluster's Jacobi
radius.  This then results in outer surface density profiles that have power law slopes in the range 
--4 to --5.  It also results in flattened velocity dispersion profiles that lie above the predictions of 
simple equilibrium 
models, with the deviation commencing at about half of the Jacobi radius \citep{KK10}.   
While \citet{KK10} caution that their results are not readily scaleable to more massive globular clusters
(recall the mass of $\omega$ Cen is a few 10$^6$ M$_\sun$), their model calculations are at least 
qualitatively in agreement with the $\omega$ Cen observations.

\begin{figure}
\centering
\includegraphics[angle=-90.,width=0.9\textwidth]{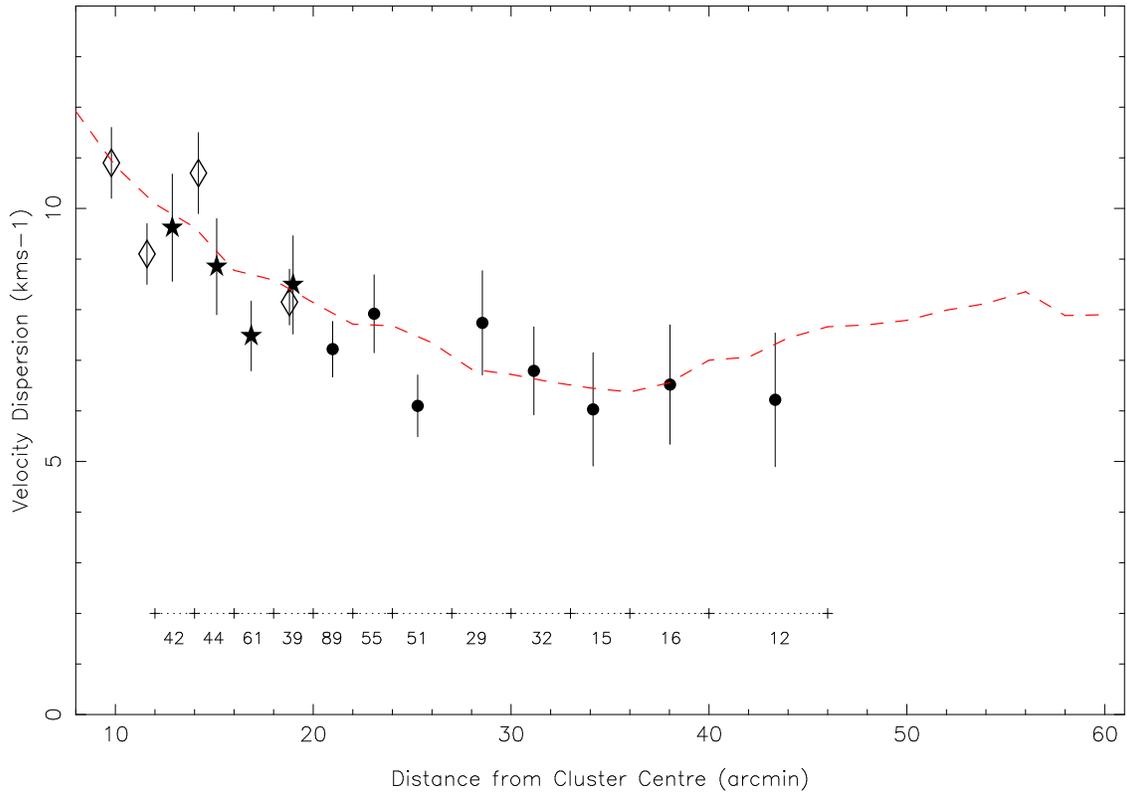}
\caption{The line-of-sight velocity dispersion of $\omega$ Cen member stars is plotted against
distance from the cluster center in arc minutes.  Symbols are as for Fig.\ \ref{paper_fig6}.
Also shown as the red dashed line is the velocity dispersion profile of the $N$-body model
described in \citet{So09} at the end of the simulation. 
\label{paper_fig9}}
\end{figure}

More specific progress in this direction is given by the results of
the N-body calculation described in \citet{So09}.  This is a $N$ = 50,000 particle model for the 
cluster, i.e., central concentration, tidal radius and mass similar to the real cluster,
calculated for an $\omega$~Cen-like orbit in a three component (bulge+disk+halo)
Galactic potential (see \citet{So09} for details).  The calculations covered $\sim$10 orbits (as against
the many 10's of orbits made by the real cluster over a Hubble time).  The
velocity dispersion profile of the bound remnant at the end of the simulation (see Fig.\ 11 of
\citet{So09}) is shown in Fig.\ \ref{paper_fig9}.  The agreement with the observations is excellent
 and it again
suggests there is no need to invoke dark matter or non-Newtonian gravity to explain the observed
velocity dispersion profile.

In summary then, the new observations presented here confirm that the velocity dispersion
profile of $\omega$~Centauri remains relatively flat at $\sim$6.5 km~s$^{-1}$ beyond approximately 
25$\arcmin$ from the cluster center.  The most likely explanation of this effect is that we are seeing 
the consequences of {\it external} influence on the dynamics of the outer parts of the stellar system, 
which contain only a small fraction of the cluster stellar mass.  Consequently, there is no {\it requirement} 
to invoke the presence of dark 
matter or non-standard gravitational theories to explain the observations.  

\acknowledgments

GDaC would like to acknowledge the contributions of Dr.\ Matthew Coleman to the initial
phases of this work, as well as the partial research support provided through Australian Research 
Council Discovery Projects grants DP0878137 and DP120101237.  Useful conversations 
with Dr.\ Agris Kalnajs and Prof.\ Ken Freeman are also gratefully acknowledged as are the
comments of the referee.  The provision of unpublished model details by Dr.\ Roeland van der
Marel and Dr.\ Antonio Sollima is also greatly appreciated, as is the excellent support at the telescope 
provided by the AAT support astronomers and night assistants.

{\it Facilities:} \facility{AAT (AAOmega)}

\end{document}